\documentclass[twocolumn,aps,prl,showpacs]{revtex4}
\usepackage{color}
\usepackage{dcolumn} 
\usepackage[dvips]{graphicx}
\usepackage{graphicx}
\newcommand{\be}{\begin{equation}}
\newcommand{\ee}{\end{equation}}

\newcommand{\bea}{\begin{eqnarray}}
\newcommand{\eea}{\end{eqnarray}}
\newcommand{\bd}{\begin{displaymath}}
\newcommand{\ed}{\end{displaymath}}
\newcommand{\ba}{\begin{array}}
\newcommand{\ea}{\end{array}}
\newcommand{\bi}{\begin{itemize}}
\newcommand{\ei}{\end{itemize}}
\newcommand{\bc}{\begin{center}}
\newcommand{\ec}{\end{center}}
\newcommand{\bfl}{\begin{flushleft}}
\newcommand{\efl}{\end{flushleft}}
\newcommand{\bfr}{\begin{flushright}}
\newcommand{\efr}{\end{flushright}}

\def\6{\partial}

\def\bra{\langle}
\def\ket{\rangle}
\def\={\!\!\!&=&\!\!\!}
\def\+{\!\!\!&&\!\!\!+~}
\def\-{\!\!\!&&\!\!\!-~}

\begin{document}
\date{\today}

\title{Bound state of 4$f$-excitation and magnetic resonance in unconventional superconductors}
\author{A. Akbari$^{1,2}$}
\author {I. Eremin$^{1,3}$}
\author {P. Thalmeier$^{4}$}
\author {P. Fulde$^{1,5}$}

\affiliation{$^1$ Max-Planck-Institut f\"ur Physik komplexer
Systeme, D-01187 Dresden, Germany \\
$^2$ Institute for Advanced Studies in Basic Sciences, P.O.Box 45195-1159, Zanjan, Iran \\
$^3$ Institute f\"ur Mathematische und Theoretische  Physik,
TU-Braunschweig,
D-38106 Braunschweig, Germany \\
$^4$Max-Planck-Institut f\"ur Chemische Physik fester Stoffe,
D-01187 Dresden, Germany \\
$^5$Asia Pacific Center for Theoretical Physics, Pohang, Korea}
\date{\today}

\begin{abstract}
We analyze the influence of unconventional superconductivity on
crystalline electric field (CEF) excitations of rare earth ions in
novel superconductors. We show that resonant magnetic excitations of
the conduction electrons that have been observed in these systems
below T$_c$ may result in the formation of the bound state in the
4$f$-electron susceptibility. This occurs at energies well below the
CEF excitation energy. The effect is discussed as a function of
temperature and the strength of the coupling between the 4$f$ and
$d$-electrons. We argue that these effects may be present in the
layered cuprates and ferropnictides which contain rare-earth ions.
\end{abstract}

\pacs{74.20.Rp, 74.25.Ha, 74.70.Tx, 74.20.-z}

\maketitle

The influence of the crystalline electric field (CEF) splitting of
rare-earth impurities on the various thermodynamic properties in
metals is well known\cite{peschel}.  It is also  known that CEF
split rare earth impurities in a conventional $s$-wave
superconductor influence the superconducting transition temperature
quite differently than non-CEF split magnetic ions described by the
Abrikosov-Gor'kov theory. There is also a feedback of conduction
electrons on the CEF splitting\cite{lowenhaupt}. In conventional
$s$-wave superconductors the Landau damping is strongly reduced for
energies smaller than twice the superconducting gap, $2\Delta_0$. In
case when the excitation energy between the ground and first excited
CEF states $\Delta_{CEF}$ is comparable to $2\Delta_0$ one finds
line narrowing of the lowest CEF excitations and a slight shift in
frequency towards lower energies when the temperature T falls below
T$_c$. This is a simple consequence of the fact that both real and
imaginary parts of the conducting electron spin response are
suppressed in an $s$-wave superconducting state for frequencies
$\omega\leq 2\Delta_0$.

The situation is different in unconventional high-T$_c$
superconductors like the layered cuprates or ferropnictides which
may contain rare-earth elements. Firstly, in these systems the
superconducting gaps are often comparable and even larger than
$\Delta_{CEF}$. This allows to study in detail the feedback effect
of superconductivity on CEF excitations. For example, partially
successful attempts have been made to analyze the symmetry of the
superconducting gap in layered cuprates by looking at the CEF
linewidth as a function of temperature with $\Delta_{CEF}\ll
2\Delta_0$\cite{furrer}. A much more dramatic effect on CEF
excitations is the strong change in the conducting electrons spin
susceptibility below T$_c$ and its consequence for the $4f$
response. In unconventional superconductors this change may result
in a resonance peak. This is a quite general phenomenon. It has been
observed by inelastic neutron scattering (INS) \cite{rossat,Osborn}
near the antiferromagnetic wave vector {\bf Q}=$(\pi,\pi)$ in a
number of systems like high-T$_c$ cuprates\cite{rossat},
UPd$_2$Al$_3$\cite{sato}, CeCoIn$_5$ \cite{broholm},
CeCu$_2$Si$_2$\cite{stockert}, and recently also in
ferropnictides\cite{Osborn}. It depends sensitively on the type of
unconventional Cooper-pairing and may also be used to eliminate
certain forms of pairing when a resonance is observed
\cite{Eremin08}.

In this letter we analyze the effect of such a resonance below the
superconducting transition temperature on CEF excitations of the
$4f$ states. We show that the formation of a resonance may cause
striking anomalies in the CEF excitation spectrum. Depending on the
strength of the coupling between the 4$f$ and $d$-electrons, the
resonance may cause the formation of the bound state (an additional
pole) in the 4$f$-electron susceptibility. We argue that these
effects may exist in the electron- and hole-doped cuprates
containing rare-earth ions. In addition we also address the
anomalous features of the CEF excitations which have been reported
in the CeFeAsO$_{0.84}$F$_{0.16}$ superconductor\cite{chi}. In the
latter case, measurements of the intrinsic linewidth and the peak
position of the lowest excited mode at $18.7$meV show a clear
anomaly at T$_c$ and an unusual temperature dependence in the
superconducting state.

The presence of rare-earth ions with $4f$ electrons in a metallic
matrix with $d$-electrons forming the conduction band can be
described by the model Hamiltonian
\bea H =  \sum_{i\gamma}\varepsilon_{\gamma}\left| i\gamma \rangle
\right. \left. \langle i \gamma \right|+\sum\limits_{{\bf k} ,\sigma
} {\varepsilon_{{\bf k}}} d_{{\bf k} \sigma }^\dag d_{{\bf k}
\sigma}+U\sum_{i,m}n_{i\uparrow}n_{i\downarrow}. \label{eq:1} \eea
Here, $\left| i\gamma \rangle \right.$ denotes  a CEF eigenstate
$\gamma$ of the incomplete 4$f$-shell at lattice site $i$, and
$d^{\dagger}_{{\bf k}\sigma }$ creates an electron in the conduction
band with wave vector ${\bf k}$ and spin  $\sigma$. The energies of
the CEF eigenstates are defined by $\varepsilon_{\gamma}$.
Furthermore, $\varepsilon_{{\bf k}}=t_{{\bf k}}-\mu$ is the
dispersion of the conduction band with $\mu$ being the chemical
potential. Correlations of the $d$-electrons are due to an on-site
electron repulsion $U$. The coupling term between the conduction and
$4f$-electrons at site $i$ is given by
\bea H_{I}=-I_{ex}(g_J-1)\sum_{i}{\bf s}_{i}{\bf
J}_i=-I_{0}\sum_{i}{\bf s}_{i}{\bf J}_i, \eea
{\it i.e.} by the exchange interaction between itinerant $d$-spins
(${\bf s}_{i}$) and localized 4$f$-electrons determined by Hund's
rule with a total angular momentum, {\bf J}. The behavior of the
conduction electron susceptibility in the superconducting state is
treated within the random phase approximation (RPA), {\it i.e.},
\begin{equation}
\chi^{(d)}_{RPA} ({\bf q}, \omega) =  \frac{\chi^{(d)}_0 ({\bf q},
\omega)}{1-U \chi^{(d)}_0 ({\bf q}, \omega)},
\end{equation}
where $\chi^{(d)}_0 ({\bf q}, \omega)$ is the non-interacting
electron susceptibility in the superconducting state. For large
momenta {\bf q} and low frequencies,  Im$\chi^{(d)}_0 ({\bf q},
\omega)$ is zero. It can exhibit a discontinuous jump at the onset
frequency of the particle-hole continuum $\Omega_c = \min \left(
|\Delta_{\bf k}| + |\Delta_{\bf k+q}| \right)$ where both {\bf k}
and {\bf k+q} lie on the Fermi surface. Note, however, that the
discontinuity in Im$\chi^{(d)}_0$ occurs only if
$\mbox{sgn}(\Delta_{\bf k}) = - \mbox{sgn}(\Delta_{\bf k+q})$ which
is not possible for an isotropic $s$-wave order parameter! A
discontinuity in Im$\chi^{(d)}_0$ leads to a logarithmic singularity
in Re$\chi^{(d)}_0$. As a result, the resonance conditions (i)
$U\mbox{Re}\chi^{(d)}_0({\bf q}, \omega_{res})= 1$ and (ii)
Im$\chi^{(d)}_0({\bf q}, \omega_{res}) = 0$ can both be fulfilled at
$\omega_{res} < \Omega_c$ for any value of $U
>0$. This results in a resonance peak below T$_c$ in form of a spin
exciton . For finite quasiparticle damping $\Gamma$, condition (i)
can only be satisfied if $U
>0$ exceeds a critical value, while condition (ii) is replaced by
U Im$\chi^{(3d)}_0({\bf q}, \omega_{res}) << 1$.

Typically a CEF splits the Hund's rule {\bf J}-multiplet of the
incomplete $4f$-shell with different CEF levels. For Ce$^{3+}$ ions
these levels are either three Kramers doublets or a doublet and a
quartet, depending on the symmetry of the CEF. We assume for
simplicity a two level system (TLS) only  consisting of two
doublets. The splitting is $\Delta_{CEF}$ and the susceptibility of
this TLS is $u_{\alpha}(\omega)=|m_{\alpha}|^2\frac{2\Delta_{CEF}}
{(\Delta_{CEF}^2-\omega^2)}$ with  $|m_{\alpha}|^2=\sum_{ij}|\bra
i|J_{\alpha}|j \ket|^2\tanh(\beta \Delta_{CEF}/2)$. For the sake of
simplicity we further assume $|m_z|\ll |m_x|$, and
$|m_x|=|m_y|=|m_{\perp}|=m_0\tanh(\beta \Delta_{CEF}/2)$ and set
$m_0 = 1$.

The 4$f$-electron susceptibility within RPA approximation is given
by
\bea \chi^{f}({\bf q},\omega) &=& \frac{u_{\alpha}({\bf q},\omega)}
{1-I_0^2u_{\alpha}({\bf q},\omega)\chi^{(d)}_{RPA}({\bf q},\omega)}
\quad. \label{eq:2} \eea
The position of the pole and its damping by
the imaginary part of $\chi^f$  can be probed by INS. It is
determined by
\bea && \Delta_{CEF}^2-\omega_{\bf q}^2-2\Delta_{CEF}
I_0^2|m_{\perp}|^2\left[\chi_{RPA}^{(d)}({\bf
q},\omega_{\bf q})\right]^{\prime}=0, \nonumber \\
&& \Gamma_q = 2
I_0^2\Delta_{CEF}|m_{\perp}|^2\left[\chi_{RPA}^{(d)}{\bf
q},\omega_{\bf q})\right]^{\prime \prime}.
\label{omega_q_eq1}
\eea

\begin{figure}[t]
\includegraphics[angle=0,width=1.0\linewidth]{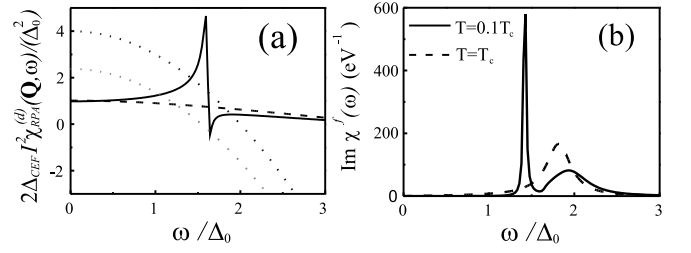}
\caption{(a) Graphical illustration of the solution of
Eq.(\ref{omega_q_eq1}). The dotted curve refers to the function
$\Delta_{CEF}^2-\omega^2$ with $\Delta_{CEF} = 2\Delta_0$ (black)
and $\Delta_{CEF}=1.5\Delta_0$ (grey), while
$2\Delta_{CEF}I_0^2\chi_{RPA}^{(d)}({\bf q},\omega_{\bf q})$ is
shown for the normal (dashed curve) and the superconducting (solid
curve) state assuming $I_0 \approx \Delta$. (b) Calculated imaginary
part of $\chi^f({\bf Q},\omega)$ at $T=T_c$(dashed curve) and in the
superconducting state ($T=0.1T_c$)(solid curve). The dispersion
parameters and Coulomb repulsion for the electron gas on a square
lattice have been used \cite{ismer}.}
\label{fig1}
\end{figure}

In the normal state Re$\chi_{RPA}^{(d)}({\bf Q}_{AF}, \omega)$ is a
positive function and decreases with increasing frequency. As a
result the actual position of the peak in Im$\chi^{f}$ will be
shifted towards lower frequency and acquire an additional damping
that originates from the Landau damping in the case of
Im$\chi^{(d)}_0$, or as in the case of Im$\chi^{(d)}_{RPA}$ by the
interaction with the overdamped spin waves of the itinerant
electrons.  This is shown in Fig.\ref{fig1}(a) where we present the
numerical solution of Eq.(\ref{omega_q_eq1}) for the normal state.
The actual position and the shift depends on the strength of the
coupling of the conduction electrons to the 4$f$-electrons.

In the superconducting state with unconventional order parameter
such that $\Delta_{\bf k} = - \Delta_{\bf k+Q}$, the conduction
electron susceptibility has a resonance peak at energy $\omega_r
\leq |\Delta_{\bf k}|+|\Delta_{\bf k+Q}|$. In this case
Re$\chi^{(d)}_{RPA}({\bf Q}, \omega)$ is discontinuous at $\omega_r$
and changes sign from positive at $\omega<\omega_r$ to the negative
at $\omega>\omega_r$. The occurrence of the resonance in the
$d$-electron susceptibility yields interesting consequences for the
4$f$-electron susceptibility. However, we have to distinguish two
cases. The feedback of the $d$-electrons on the $4f$-susceptibility
will be different in the case when the interaction between both is
strong or weak.

In the first case determined by condition $I_0^2 \geq
\frac{\Delta_{CEF}^2-\omega_r^2}{\Delta_{CEF}
\mbox{Re}\chi^{(d)}_{RPA}({\bf Q}, \omega_r)}$,  the 4$f$-electron
susceptibility will have an additional pole (bound
state\cite{thalm}) at energies well below $\Delta_{CEF}$.
Simultaneously, due to the change of sign of the
$\chi^{(d)}_{RPA}({\bf q}, \omega)$ for $\omega > \omega_r$, the
initial position of the $\Delta_{CEF}$ will be shifted to higher
frequencies. The resulting pole structure and Im$\chi^f({\bf
Q},\omega)$ are shown in Fig.\ref{fig1}. In other words, due to the
interaction between the 4$f$-shell and $d$-electrons and assuming $
\Delta_{CEF}\simeq 2\Delta_0$, the positions of the new poles are
approximately given by
%
$\omega_q^{\pm} \simeq  \frac{1}{2}(\Delta_{CEF}+ \omega_r)
\pm \frac{1}{2}\sqrt{\left(\Delta_{CEF}-
\omega_r\right)^2+4I_0^2|m_{\perp}|^2z_r}$
%
where $z_r$ is residue of the resonance peak in the $d$ spin
susceptibility.
\begin{figure}[t]
\includegraphics[angle=0,width=1.0\linewidth]{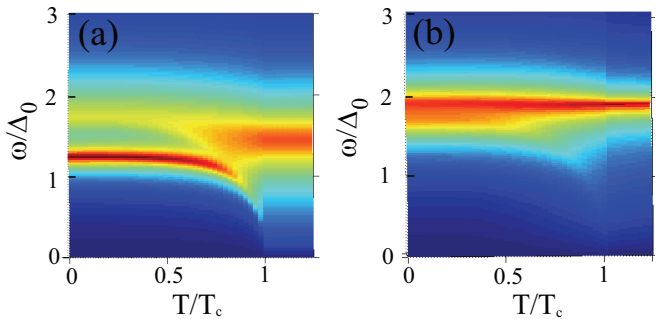}
\caption{Calculated temperature dependence of the lowest CEF
excitation in an unconventional superconductor for (a) strong
coupling ($I_0 \approx \Delta_0$) and (b) weak coupling ($I_0 <
\Delta_0$) coupling between conduction electrons and localized
4$f$-electrons. The tight-binding parameters for the electron-doped
cuprates\cite{ismer} and ferropnictides\cite{Korshunov08} are used.
In both cases we assume a temperature dependence of the
superconducting order parameter of the form $\Delta_{{\bf
k}}(T)=\Delta_{\bf k}\tanh\left(1.76\sqrt{\frac{T_c}{T}-1}\right)$.
}
\label{fig2}
\end{figure}

In Fig.\ref{fig1}(b) we show the total structure of the imaginary
part of the $f$-electron spin susceptibility as function of
frequency at the antiferromagnetic wave vector, {\bf Q}. Due to
appearance of the resonance mode in the conduction electron
susceptibility the CEF excitations acquire an additional pole below
the superconducting transition temperature at energies well below
$2\Delta_0$ with a small linewidth. In addition a second pole is
found at a frequency which is higher than the renormalized
$\Delta_{CEF}$ in the normal state. At the same time damping induced
by Im$\chi_{RPA}^{(d)}$ is still larger than it is in the normal
state value as a consequence of discontinuous jump of Im$\chi_0$ at
$2\Delta_0$ induced by the requirement $\Delta_{\bf k} = -
\Delta_{\bf k+Q}$. Therefore, CEF excitations behave completely
opposite as in conventional superconductors. It is also interesting
to note that in case $\Delta_{CEF}$ becomes smaller than $2\Delta_0$
the position of the bound state shifts to lower energies (see
Fig.\ref{fig1}(a)). Moreover, at certain value of the $\Delta_{CEF}<
1.5\Delta_0$ one finds a single pole as a result of the coupling
between $d$ and 4$f$-susceptibilities representing the renormalized
CEF excitations. Furthermore, the second pole related to the bound
state no longer exists. In Fig.\ref{fig2}(a) we show the temperature
evolution of the CEF excitations below the superconducting
transition temperature. Here we assume a tight-binding energy
dispersion for the electron-doped cuprates and a $d$-wave symmetry
for the superconducting gap, {\it i.e.} $\Delta_{\bf k} =
\frac{\Delta_0}{2}\left(\cos k_x - \cos k_y \right)$ \cite{ismer}.
Using a temperature dependence for the superconducting gap that
resembles the solution obtained for the Eliashberg strong-coupling
equations ($\Delta_{{\bf k}}(T)=\Delta_{\bf k}
\tanh\left[1.76\sqrt{\frac{T_c}{T}-1}\right]$) we analyze the CEF
excitations below the superconducting transition temperature. Close
to T$_c$, {\it i.e.}, in the range 0.8T$_c < T < T_c$, the splitting
of the CEF excitations due to formation of the magnetic resonance is
not well resolved and it looks as if the CEF level gets damped
anomalously strong. With lowering temperature,{\it i.e.},
T$\approx$0.8T$_c$, both peaks become well separated. At even lower
temperatures both peaks are clearly separated in energy and the
lower mode is much narrower.

In the second case when the coupling between conduction electrons an
localized 4$f$-electrons is weak, the effect of unconventional
superconductivity on the CEF excitations still differs substantially
from that of conventional superconductors. In particular, for $I_0 <
\Delta_0$ there is no additional peak in Im$\chi^f$ below T$_c$.
This is due to relatively small value of $I_0^2\chi_{RPA}^{(d)}({\bf
q},\omega_{\bf q})$. This is also the case if the damping of the
magnetic resonance in $\chi^{(d)}$ is relatively strong. In other
words, even though $\chi^{(d)}$ is strongly frequency dependent at
low energies the coupling is not large enough to produce a second
pole in the $f$-electron susceptibility. Nonetheless, the effect of
the resonance in the former is present in this frequency range.
Because the overall conduction electrons response of the
unconventional superconductor is larger than its normal state
counterpart, the CEF excitations below T$_c$ will experience larger
damping due to Im$\chi_{RPA}^{(d)}({\bf Q}, \omega \sim 2\Delta_0)$
which is enhanced in the superconducting state. Similarly
Im$\chi_{RPA}^{(d)}({\bf Q}, \omega \sim 2\Delta_0)$ is lower than
its normal state value but, more important, it changes sign above
$\omega_r$ (see Fig.\ref{fig1}(a)). As a result the CEF excitation
simultaneously becomes more damped and also shifts towards higher
frequencies. We show the evolution of the CEF excitation in
Fig.\ref{fig2}(b) as a function of temperature where below this
effect is clearly visible as a function of temperature.

The description above is correct under the assumption that there is
a well-defined resonant excitation in the conduction electron
susceptibility at the two-dimensional antiferromagnetic wave vector,
{\bf Q}. This implies that for all points of the Fermi surface the
condition $\Delta_{\bf k} = - \Delta_{\bf k+Q}$ is satisfied. It has
been shown previously, that this condition is realized particularly
well in quasi-two dimensional systems such as hole- and
electron-doped cuprates\cite{ismer} as well as in the novel
iron-based superconductors\cite{Korshunov08} and also in other
heavy-fermion systems like CeCoIn$_5$\cite{Eremin08}.

\begin{figure}[t]
\includegraphics[angle=0,width=1.0\linewidth]{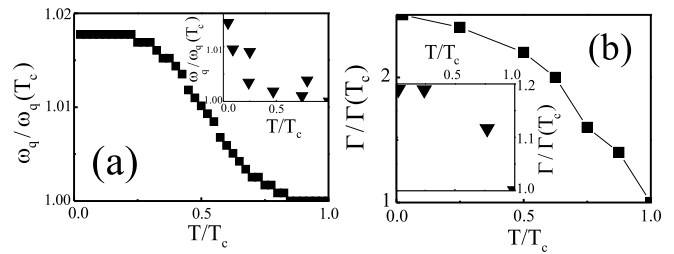}
\caption{Calculated temperature dependence of the frequency
position, $\omega_q$ (a) and the linewidth (b) of the lowest CEF
level assuming weak coupling between the conduction and the
4$f$-shell. The insets show the corresponding experimental data for
the superconducting CeFeAsOF system\cite{chi}. }
\label{fig3}
\end{figure}

It has been found recently that lowest CEF excitations of Ce ions
located at around $2\Delta$ shows an anomalous behavior in the
superconducting state of CeFeAsO$_{1-x}$F$_x$, see Ref.
\onlinecite{chi}. In particular, the lowest CEF excitations shifts
slightly towards higher energies and acquire an additional damping
below T$_c$. Assuming the extended $s$-wave symmetry of the
superconducting gap that most likely is realized in ferropnictide
superconductors\cite{mazin}
($\Delta_{k}=\frac{\Delta_0}{2}\left(\cos k_x +\cos k_y\right)$),
this unusual behavior can be well understood. In particular, if the
coupling between the conduction electrons and Ce ions is relatively
weak ($I_0 < \Delta_0$) the effect of the unconventional
superconductivity on the CEF excitations does not result in the
appearance of an additional pole in Im$\chi_f$, since the coupling
is too weak. However, because of the sign change of the
superconducting gap for the electron and the hole Fermi surfaces
connected by ${\bf Q}_{AF}$, Im$\chi_{RPA}^{(d)}$ enhances in the
superconducting state at energies $\omega>\omega_r$. Simultaneously,
the real part of $\chi_f$ changes sign and becomes negative for
$\omega>\omega_r$ which leads to the negative shift of
$\Delta_{CEF}$ below T$_c$, {\it i.e.} it shifts to higher energies
below T$_c$(see Fig.\ref{fig2}(b)). In order to compare our results
to the experimental data we show in Fig.\ref{fig3} the calculated
shift of the lowest frequency of the CEF excitations as well as
their damping (half width at half maximum) in the superconducting
state normalized to their values at T$_c$. We adopt the parameters
used previously for calculations of the itinerant magnetic
excitations in ferropnictides\cite{Korshunov08}. We further assume
the coupling of the itinerant $3d$ electrons to the 4$f$-shell
$I=4.3$meV$< \Delta$. In the superconducting state, due to the
extended $s$-wave symmetry of the superconducting gap the itinerant
interband RPA spin susceptibility between electron and hole Fermi
surface pockets centered around the $\Gamma$ and $M$ points of the
BZ, respectively, shows the characteristic enhancement at energies
smaller than $2\Delta$\cite{Korshunov08,maier}. Due to weak coupling
to the 4$f$-shell this does not yield a second pole in the
Im$\chi_f$ but results in the anomalous damping of the CEF
excitations and a shift of the characteristic frequency, $\omega_q$
towards higher energies. We find that this behavior qualitatively
and also to large extent quantitatively agrees with the experimental
data on CeFeAsO$_{0.84}$O$_{0.16}$\cite{chi}.

The other systems where similar effects can realize but with strong
coupling between the conduction and 4$f$-electrons are layered
cuprates with rare-earth ions or heavy-fermion superconductor
CeCoIn$_5$. For example, in the electron-doped cuprate system
Nd$_{2-x}$Ce$_x$CuO$_4$, the splitting between the ground and the
first CEF excited levels of the 4$f$ multiplet for Nd$^{3+}$ ions is
about $\Delta_{CEF} \approx$20meV\cite{boothroyd} which is
comparable to the energy of 2$\Delta$ of the superconducting gap in
the same system\cite{ismer}. At the same time, the situation with
the position of the feedback spin resonance due to the
$d_{x^2-y^2}$-wave symmetry of the superconducting gap in the
electron-doped cuprates is far from being well understood. For
example, originally the resonance has been reported in
Pr$_{0.88}$LaCe$_{0.12}$CuO$_{4-\delta}$ and
Nd$_{1.85}$Ce$_{0.15}$CuO$_{4}$ at the energies of about 12meV and
10meV, respectively\cite{wilson,zhao}. Recently, this conclusion has
been challenged by another group\cite{greven} where the feedback of
superconductivity in Nd$_{1.85}$Ce$_{0.15}$CuO$_{4}$ has been found
at much smaller energies around $4.5$meV and $6.4$meV. In
Fig.\ref{fig1}(b) we show the calculations for Im$\chi_f$ assuming a
tight-binding energy dispersion for the 3$d$-electrons \cite{ismer},
the coupling constant $I_0 = \Delta_0$ and $\Delta_{CEF} =
2\Delta_0$. One finds that in this case the $f$-electron
susceptibility shows two peaks below $T_c$ with the lower one at the
energy smaller than $\omega_r$ and the upper one slightly above the
renormalized $\Delta_{CEF}$ in the normal state. In order to see
whether the lower peak found recently\cite{greven} is indeed related
to the feedback effect of the unconventional superconducting order
parameter on the CEF excitations further studies are necessary. A
hallmark of this would be the large contribution of the incomplete
4$f$-shell to the $d$-electron susceptibility spin resonance. In
addition, below T$_c$ the CEF excitations should possess larger
damping in the superconducting state than in the normal state and
show slight shift towards higher energies.

In conclusion, we find that the feedback of unconventional
superconductivity on the CEF excitations may results in two
characteristic features. If the coupling between the CEF excitations
and conduction $d$-electrons is large ($I_0\sim \Delta$), the
resonant excitations in the conducting electrons susceptibility
centered at $\omega_r$ yield an additional bound state in the
$f$-electron susceptibility at energies $\omega_1\leq \omega_r$. At
the same time, the CEF excitations shifts towards higher energies
and acquire an additional damping below T$_c$. If the coupling
between the $d$-electrons and CEF excitations is weak, {\it i.e.}
$I_0<\Delta$ the additional pole does not occur and the only effect
of the unconventional superconductivity is the anomalous damping of
the CEF excitations and their slight upward shift below T$_c$. We
argue that both these effects may be present in different layered
cuprates and ferropnictides, respectively.

We would like to thank M.M. Korshunov, P. Dai, and M. Greven for
useful discussions. I.E. and P.T. acknowledge support from the Asia
Pacific Center for Theoretical Physics where part of this work has
been done. I.E. is partially supported by the Volkswagen Foundation
(I/82203), NSF-DMR (0645461), and the RAS Program (N1 2.1.1/2985).


\begin{thebibliography}{99}

\bibitem{peschel} P. Fulde and I. Peschel, Adv. Phys. {\bf 21}, 1
(1972).
\bibitem{lowenhaupt} J. Keller, and P. Fulde, J. Low Temp. Phys. {\bf
12}, 63 (1973).
%
\bibitem{furrer} J. Mesot, G. B\"ottger, H. Mutka, and A. Furrer,
Europhys. Lett. {\bf 44}, 498 (1998).
%
\bibitem{rossat} J. Rossat-Mignod {\it et al.},
Physica C {\bf 185-189}, 86 (1991).
%
\bibitem{Osborn} A.D. Christianson {\it et al.},
Nature {\bf 456}, 930 (2008).
%
\bibitem{eschrig} For a review see M. Eschrig, Adv. Phys. {\bf 55}, 47 (2006).

\bibitem{sato} N. K. Sato {\it et al.},
Nature (London) {\bf 410} 340 (2001).

\bibitem{broholm} C. Stock, C. Broholm, J. Hudis, H J. Kang, and C.
Petrovic, Phys. Rev. Lett. {\bf 100}, 087001 (2008).

\bibitem{stockert} O. Stockert {\it et al.}, Physica (Amsterdam) 403B, 973 (2008).

\bibitem{Eremin08} I. Eremin, G. Zwicknagl, P. Thalmeier and P. Fulde, Phys. Rev. Lett. {\bf 101}, 187001
(2008).

\bibitem{chi} S. Chi {\it et al.},
Phys. Rev. Lett. {\bf 101}, 217002 (2008).


\bibitem{Korshunov08}M.M.~Korshunov and I.~Eremin, Phys. Rev. B {\bf 78}, 140509(R) (2008)
%
\bibitem{ismer} J.-P. Ismer, I. Eremin, E. Rossi, and D.K. Morr, Phys. Rev. Lett. {\bf 99}, 047005
(2007).

\bibitem{thalm} A bound state between a phonon and CEF excitation is discussed in P. Thalmeier, and
P. Fulde, Phys. Rev. Lett. {\bf 49}, 1588 (1982).

\bibitem{boothroyd} A.T. Boothroyd, S.M. Doyle, D.M. Paul,
and R. Osborn, Phys. Rev. B {\bf 45}, 10075 (1992).
%
\bibitem{wilson} S.D. Wilson, P. Dai, Sh. Li, S. Chi, H.J. Kang, and J.W.
Lynn, Nature {\bf 442}, 59 (2006).

\bibitem{zhao} J. Zhao, P. Dai, S. Li, P.G. Freeman, Y. Onose, and Y. Tokura
Phys. Rev. Lett. {\bf 99}, 017001 (2007).

\bibitem{greven} G. Yu {\it et al.},
arXiv:0803.3250 (unpublished).

%
%
\bibitem{mazin}  I.I. Mazin, and J. Schmalian, arXiv:0901.4790
(unpublished).
%
\bibitem{maier} T.A. Maier and D.J. Scalapino, Phys. Rev. B {\bf 78},
020514(R) (2008).

%
\end{thebibliography}
\end{document}